\documentclass[main]{sshArticle}

\usepackage{stfloats} 
\usepackage{nicefrac} 
\definecolor{colorhbar}{RGB}{154, 181, 255}

\usepackage{amsmath}
\usepackage{amssymb}
\usepackage{graphicx}
\usepackage{gensymb}
\usepackage{textcomp}
\usepackage{bm}
\usepackage{placeins}
\usepackage{soul}
\usepackage{xcolor}

\title{Ultra-Cold Cryogenic TEM with Liquid Helium and High Stability}

\author[1,2,*]{Emily Rennich}
\author[1,3,*]{Suk Hyun Sung}
\author[3]{Nishkarsh Agarwal}
\author[2]{Maya Gates}
\author[4]{Robert Kerns}
\author[3,5,†]{Robert Hovden}
\author[1,†]{Ismail El Baggari}

\affil[1]{The Rowland Institute at Harvard, Harvard University, Cambridge, MA 02142, United States}
\affil[2]{Department of Mechanical Engineering, University of Michigan, Ann Arbor, MI 48109, United States}
\affil[3]{Department of Materials Science and Engineering, University of Michigan, Ann Arbor, MI 48109, United States}
\affil[4]{Michigan Center for Materials Characterization, University of Michigan, Ann Arbor, MI 48109, United States}
\affil[5]{Applied Physics Program, University of Michigan, Ann Arbor, MI 48109, United States}

\affil[*]{Contributed equally}
\affil[†]{e-mail: hovden@umich.edu, ielbaggari@rowland.harvard.edu}

\date{\today}

\firstAuthorLastName{Rennich}

\abstract{Cryogenic transmission electron microscopy has revolutionized structural biology and materials science, but achieving temperatures below the boiling point of liquid nitrogen remains a long-standing aspiration.
We introduce an ultra-cold liquid helium transmission electron microscope specimen holder, featuring continuous cryogen flow and vibration decoupling.
This instrument is compatible with modern aberration-corrected microscopes and achieves sub-25~K base temperature, ±2~mK thermal stability over many hours, and atomic resolution---setting the stage for a new era of cryogenic electron microscopy.
}

\begin{document}

\maketitle
Rapid advancements of cryogenic transmission electron microscopy (TEM) techniques have revolutionized the biological~\cite{kuhlbrandt2014resolution,yip2020atomic} and materials sciences~\cite{li2017atomic,zachman2018cryo,elbaggari2018,zhu2023formation} over the past decades.
However, serious challenges persist in the realm of cryogenic TEM techniques, particularly when working at temperatures below 100~K. 
Ultra-cold temperatures below the boiling point of liquid nitrogen (77 K) in modern TEM are poised to improve the radiation tolerance and resolution in single-particle cryo-EM and cryo-electron tomography~\cite{henderson2004realizing,dubochet1981low,naydenova2022reduction,russo2022cryomicroscopy}, and provide access to quantum phases in materials science research~\cite{harada1992real,zhao2018direct,minor2019cryogenic,zhu2021cryogenic,naydenova2020cryo}.

Dedicated helium stage modules were constructed and implemented early in the history of TEM~\cite{venables1963liquid,matricardi1967electron,fujiyoshi1991development,jiang2008backbone,borrnert2019dresden}; 
however, this approach is currently incompatible with high-resolution objective lens pole pieces due to the strict design constraints of modern electron optics.
In life science applications, cartridge-based liquid nitrogen microscopes have become popular even though they lack ultra-cold performance and in situ capabilities. 
As an alternative, dewar-based side-entry cryogenic holders have filled the need to cool specimens to low temperatures.
Unfortunately, the rapid evaporation of the cryogen within the dewar introduces large mechanical vibrations and thermal instabilities that result in poor imaging resolution. 
Moreover, the relatively small volume of these dewars leads to short operation times, or hold times~\cite{minor2019cryogenic,zhu2021cryogenic}. 
While some of these challenges can be mitigated for liquid nitrogen, liquid helium has a heat of vaporization $\sim$60 times lower than that of liquid nitrogen, which dramatically increases instabilities and has made the use of liquid helium a persistent challenge for cryogenic TEM~\cite{zhu2021cryogenic}.

Here we introduce a side-entry specimen holder capable of liquid helium cooling while achieving atomic resolution. 
This ultra-cold cryogenic TEM specimen holder demonstrates millikelvin temperature stability maintained for hours near liquid helium temperatures ($<$ 25~K at the specimen) in a modern aberration-corrected TEM. 
For the first time, a base temperature of 23~K is maintained over 4 hours. 
Moreover, the thermal stability is better than $\pm$2~mK, a significant (estimated 10$\times$) improvement over existing instruments. 
Thermal stability is directly associated with the reduction of specimen drift. 
By maintaining stable cryogenic temperatures with liquid helium for extended durations, scientists can conduct intricate experiments without grappling with unpredictable temperature fluctuations or recurrent dewar refills.

\begin{figure*}[!t]
\centering
  \includegraphics[width=\textwidth]{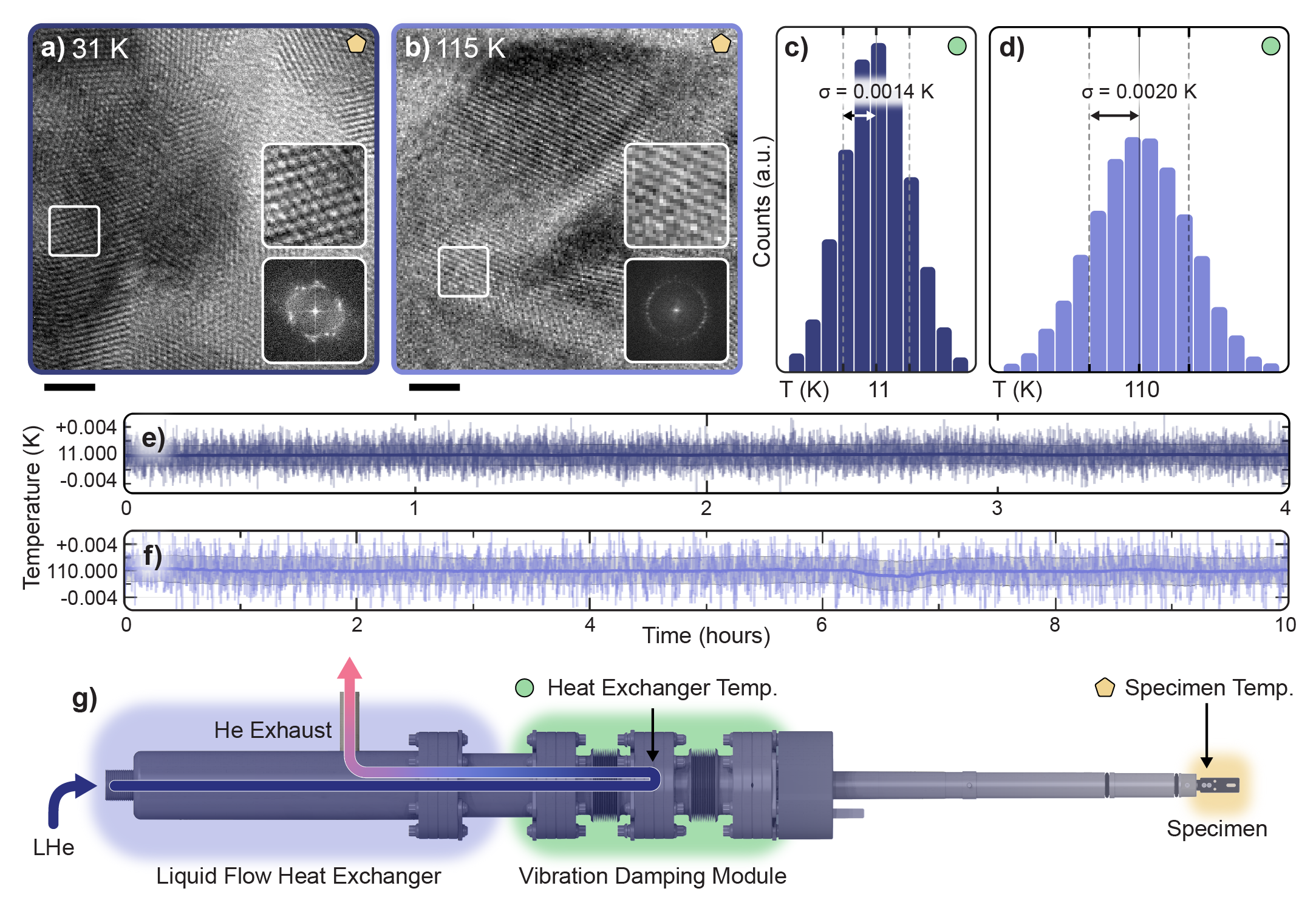}
  \caption{\textbf{Ultra-cold cryogenic TEM imaging at atomic resolution with mK temperature stability.} a),b Atomic resolution image of gold at 31~K, and 115~K, respectively. 
  The top insets represent a zoom-in into the TEM images.
  Bottom insets show the Fourier transform of the atomically resolved TEM images.
  The scale bar represents 2 nm. 
  c,d) Histograms of measured sensor temperatures at 11~K and 110~K, respectively.
  A temperature stability of ±1.4 mK was measured at a heat exchanger base temperature of 11~K over a data collection period of 4 hours. 
  A temperature stability of ±2.0 mK was measured at an intermediate heat exchanger temperature of 110~K over a data collection period of 4 hours. 
  e),f) Plot of the measured heat exchanger temperature fluctuations as a function of time at 11 K and 100 K, respectively.
  Multi-hour long hold times with $\leq$2 mK stability were demonstrated at both temperatures. 
  g) Rendered model of the instrument components including the cryogen-flow heat exchanger, the vibration damping module, and TEM specimen tip.
  Liquid helium flow through the heat exchanger is denoted using an arrow with relative holder temperature denoted by a blue-white gradient overlay. 
  Temperature measurements were taken using a calibrated silicon diode sensor at the specimen tip and liquid flow heat exchanger. Temperature difference between heat exchanger and specimen tip is $\sim$15~K.
  }
  \label{F:Fig1}
\end{figure*}

Figure~\ref{F:Fig1}a shows the atomic structure of gold at $\sim$31~K using this ultra-cold TEM specimen holder inserted into a double-corrected JEOL 3100R05 operating at 300 kV. 
This holder (Fig.~\ref{F:Fig1}g) employs liquid cryogen flow through a copper heat exchanger as the cooling source. 
Liquid helium continuously flows from a large external dewar along a vacuum-insulated and radiation-shielded transfer line to the copper heat exchanger. 
The sample is coupled to the cooled heat exchanger via highly thermally conductive components running axially inside the long hollow metal rod of the holder, keeping it cold as long as liquid helium is flowing.
The extended surface area of these cold components act as internal cold traps in the vacuum space, preventing ice accumulation on the specimen even after multiple hours of experiments.

Vibrations are mitigated by decoupling the specimen from the helium transfer line through two pairs of flexible ultra-high vacuum edge-welded bellows with a spring constant of 4.7~N/mm.
Each bellow compresses 50A durometer rubber blocks that dampen vibrations emanating from the transfer line and room environment. 
The flexible bellows also accommodate movements of the microscope's goniometer, enabling the usual positional adjustments in the $x$, $y$ and $z$ axes, as well as axial rotation. 

In addition to low base temperatures, this liquid helium cryogenic TEM holder allows precise control over a wide range of temperatures without compromising thermal stability. 
By adjusting the flow of helium and applying carefully controlled heating, the holder can be stabilized anywhere between room and base temperature with a stability of $\pm$2~mK or better. 
This capability has been demonstrated, reaching $\pm$1.4~mK and $\pm$2.0~mK stability at temperatures of 11~K and 110~K, respectively (see Fig.~\ref{F:Fig1}c-f). 
This instrument is also cryogen-agnostic: liquid nitrogen can be used instead of liquid helium if only temperatures above 110~K are needed.

Millikelvin stability over 10 hours is shown in the temperature plot (Fig.~\ref{F:Fig1}f) at an intermediate temperature of 110~K, though similarly long hold times are possible at any operating temperature. 
In fact, hold times are only limited by the available helium supply---this holder consumes helium at a rate of less than two litres per hour.
Commonly available helium tanks are between 30--1000 liters, making long hold times easily achievable.
These long hold times present a substantial opportunity for measurements that require long experimental durations including single particle reconstruction, chemical spectroscopy, tomography, and ptychography. 
The ability to easily reach intermediate temperatures also affords a great deal of flexibility: materials can now be reliably characterized at any desired temperature, not just near the boiling point of liquid nitrogen.

\begin{figure*}[!t]
\centering
  \includegraphics[width=\textwidth]{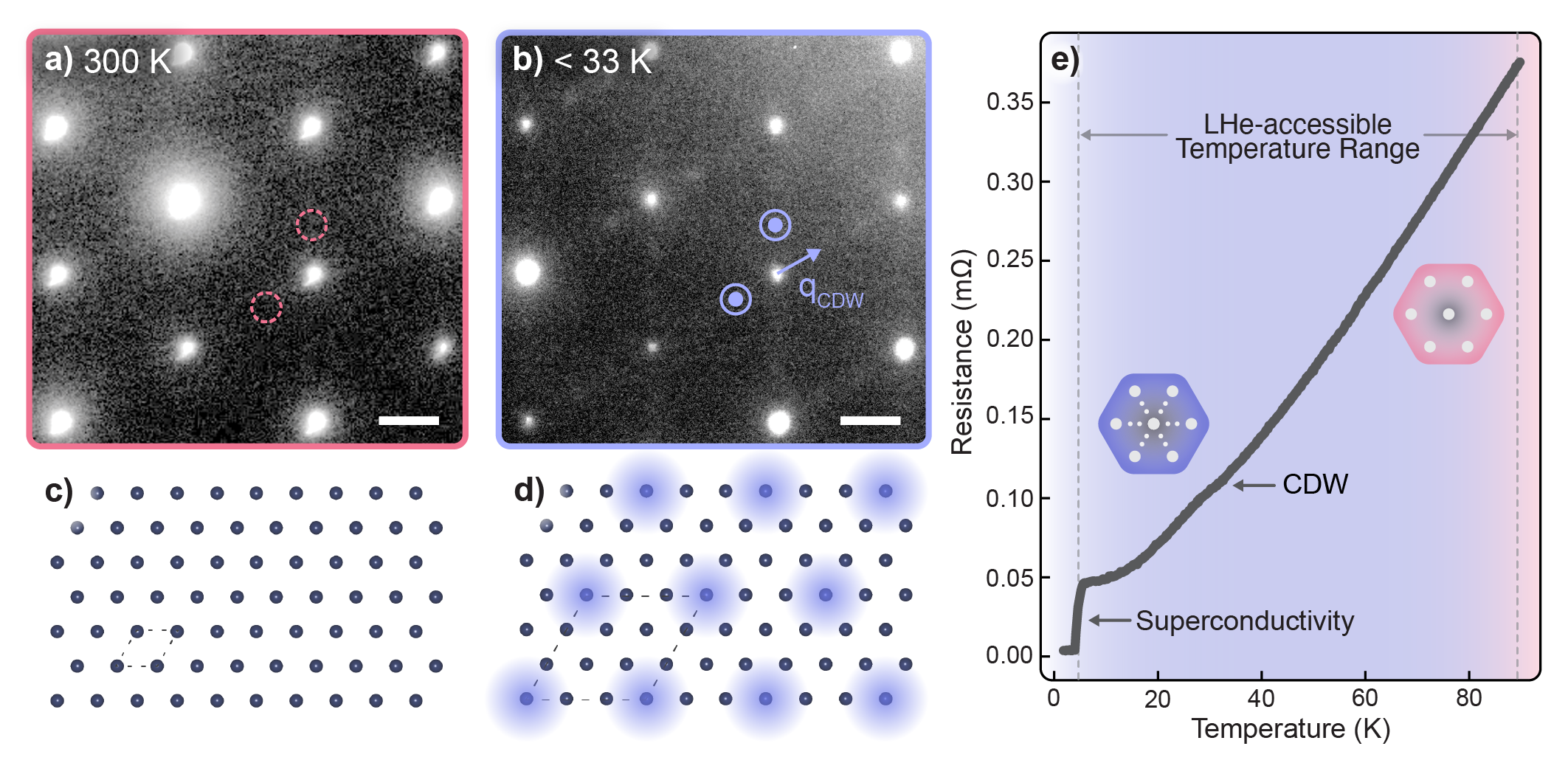}
  \caption{\textbf{The emergence of a charge density wave superlattice in 2H-NbSe\textsubscript{2} below 33~K.} 
  a) TEM electron diffraction pattern captured at 300~K shows the room temperature phase containing only Bragg peaks corresponding to a c) normal crystal lattice. 
  b) At temperatures below 33~K superlattice peaks (blue circles) appear, indicating emergence of charge density waves (denoted by blue intensities) with a threefold supercell in d). 
  The 1/3 reciprocal lattice unit wavevector is represented by a blue arrow.
  c) Simplified model of the hexagonal atomic arrangement of Nb atoms (dark circles) in 2H-NbSe\textsubscript{2}. 
  Dashed lines represent the unit cell at high temperature.
  The lattice constant representing the Nb-Nb distance is 0.34~nm. 
  d) Model of the atomic arrangement overlaid of the charge density wave superlattice (dashed lines).
  The light blue circles represent the extra charge density that forms a supercell at low temperature.
  e) Temperature-dependent resistance curve shows the emergence of electronic phase transitions including a CDW transition at $\sim$33 K and superconductivity below 7 K.
  The CDW transition appears as a slight anomaly in the temperature-resistance curve whereas the superconducting transition appears as a sudden drop to zero resistance.  
  The scale bars represent 2~nm\textsuperscript{-1}. 
  }
  \label{F:FigL}
\end{figure*}

Beyond applications for the preservation of beam-sensitive specimens, ultra-cold TEM opens up avenues for investigating quantum materials, crystalline solids in which novel electronic states emerge at low temperatures. 
An illustrative example is the observation of a structural phase transition below 33~K in the layered quantum material 2H-NbSe\textsubscript{2} (Fig.~\ref{F:FigL}), which further validates ultra-cold TEM performance. 
2H-NbSe\textsubscript{2} is a layered transition-metal dichalcogenide that exhibits charge density wave (CDW) ordering below 33~K and superconductivity below 7~K~\cite{ugeda2016characterization}. 
CDW formation is characterized by the appearance of superlattice peaks in reciprocal space.
Figure~\ref{F:FigL}a shows an electron diffraction pattern at 300~K from a thin exfoliated flake. 
Bragg peaks with the hexagonal symmetry of the crystal are prominent. 
Accessing temperatures below 33~K with the ultra-cold holder, we observe the emergence of additional superlattice peaks (Fig.~\ref{F:FigL}b, circles) indicating the formation of the CDW. 
These peaks are located at $\nicefrac{1}{3}$ reciprocal lattice units, the predicted wavevector for CDW in 2H-NbSe\textsubscript{2}. 

The results herein represent a significant advancement that addresses a longstanding desire to access ultra-cold temperatures within a modern aberration-corrected electron microscope, offering new opportunities for probing challenging biological and quantum materials.
Cryogenic electron microscopy at ultra-cold temperatures is critical to improving radiation tolerance and imaging resolution of key materials underlying biological and renewable energy research.
This development also enables a new field for characterization of quantum materials, including mapping of phase transitions such as superconductivity at temperatures unattainable with liquid nitrogen cooling.

\section*{References}
\par\vspace{5pt}
{\printbibliography[heading=none]}

\section*{Methods}

\subsection*{Cryogenic Experiment}
The holder incorporates liquid helium flow cooling and flexible thermal and vacuum connections to reduce mechanical vibration coupling. 
The internal cryogenic components operate under high vacuum conditions provided by the microscope column vacuum ($\sim$1.0$\times$10\textsuperscript{-7} torr).
Due to the increase in internal surface area, the holder is pumped under vacuum for at least 30 minutes before imaging.
Internally, a copper heat exchanger is connected to a copper braid and rod assembly that traverses the bore of the TEM holder.

A liquid helium transfer line connects the specimen holder to a large external helium source---here a 60~L stainless steel dewar was used. 
The dewar is depressurized when the helium transfer line is inserted and secured to the dewar. 
Pressure in the dewar is then raised and maintained at  $\sim$5~psi to ensure flow into the ultra-cold specimen holder.

For precise temperature control, we use a flexible heating element attached to the heat exchanger to maintain the desired temperature.
The temperature is measured using a silicon diode with an excitation DC current of 10~μA and controlled using a temperature controller (Cryo-Con Model 22C) with a proportional-integral-derivative (PID) scheme. 
A second temperature sensor is mounted at the tip, near specimen to monitor specimen temperature.

\subsection*{Specimen Preparation}
Sputtered gold on an ultrathin carbon film covered a standard 3 mm copper mesh grid. 
2H-NbSe\textsubscript{2} was exfoliated onto polydimethylsiloxane (PDMS) gel stamp and mechanically transferred using a home-built transfer system onto silicon-based TEM grids. 
The silicon based TEM grids contain a SiN\textsubscript{x} membrane window TEM grid with 2~μm holes and were produced by Norcada.

\subsection*{Temperature-Dependent Resistance Measurements}
Temperature-dependent resistance measurements on 2H-NbSe\textsubscript{2} were performed in a 1.8~K cryostat with a superconducting magnet (Quantum Design PPMS DynaCool).
Samples were prepared by exposing a fresh surface of 2H-NbSe\textsubscript{2} by scotch-tape exfoliation and thin gold wires were attached using silver paint. 
Resistance measurements were done in a 4-probe configuration with a 1~mA current being sourced from a Keithley 6220 Current Source and voltage measured via a Keithley 2812A Nanovoltmeter. 
The sample was cooled in helium exchange gas.
The measurements shown here were performed under a 2~T out-of-plane magnetic field to match the magnetic field experienced by the sample in a TEM.
Due to this magnetic field, the superconducting transition is suppressed to $\sim$5~K instead of the $\sim$7~K transition temperature under ambient conditions.
The charge density wave transition is unaffected by the magnetic field.

\subsection*{Transmission Electron Microscopy}
The in situ cooldown was performed on a double Cs-Corrected JEOL3100R05 STEM/TEM equipped with tungsten cold-FEG operated at 300~kV held under a $\sim$5.2$\times$10\textsuperscript{-11} torr vacuum. 
The microscope column pressure of $\sim$7.5$\times$10\textsuperscript{-8} torr was maintained around the specimen throughout the experiment. 
TEM images (Fig.~\ref{F:Fig1}a,b) were acquired on Gatan Ultrascan1000 CCD camera with 25~ms acquisition time. 
Electron diffraction patterns were collected using an 850~nm selected area aperture.

\section*{Acknowledgements}
The first liquid helium cool-down occurred on July 1, 2022. Sub-30~K CDW phase transition 2H-NbSe$_2$ with hold times in excess of ten hours was observed on August 3, 2022. Atomic resolution images at 23~K where acquired on July 18, 2023.
This research was supported by the Rowland Institute at Harvard.
We acknowledge key contributions from Winfield Hill, Erik Madsen, Kal Banger, Alan Stern and Chris Stokes from the Rowland Institute at Harvard for their help with machining, electronics design, and cryogenics. 
Electron microscopy experiments were conducted using the Michigan Center for Materials Characterization ((MC)\textsuperscript{2}) at the University of Michigan with assistance from Tao Ma, Haiping Sun, and Allen Hunter. 
I.E. and E.R. acknowledge support from the Rowland Institute at Harvard. 
R.H. acknowledges support from the U.S. Department of Energy, Basic Energy Sciences, under award DE-SC0024147. 
E.R. acknowledges support from the National Science Foundation award 2228909.

\section*{Author contributions}
I.E., R.H., E.R. conceived the design and experiments. 
E.R., S.H.S., N.A., M.G., R.H, and I.E. performed cooldown experiments and in situ microscopy measurements. 
I.E., S.H.S., N.A., M.G., E.R., R.K., and R.H. built the experimental microscopy setup and performed hardware assembly. 
S.H.S. and N.A. optimized temperature control.
E.R., S.H.S., R.H., and I.E. prepared the manuscript. 
All authors reviewed and edited the manuscript.

\section*{Competing interests}
E.R., R.H., I.E. are inventors on a filed patent application related to this work and licensed for commercialization. 
The other authors declare no competing interests.

\end{document}